\input phyzzx.tex
\tolerance=1000
\voffset=-0.0cm
\hoffset=0.7cm
\sequentialequations
\def\rl{\rightline}

\def\t1{{\tilde 1}}

\def\t{\theta}

\REF{\STR}{J. M. Maldacena, [arXiv:hep-th/9607235],
A. W. Peet, [arXiv:hep-th/0008241]; J. R. David, G. Mandal and S. R. Wadia, Phys. Rep. {\bf 369} (2002) 549, [arXiv:hep-th/0203048].}
\REF{\LAST}{E. Halyo, [arXiv:1502.01979], [arXiv:1503.07808].}
\REF{\LEN}{L. Susskind, [arXiv:hep-th/9309145].}
\REF{\SBH}{E. Halyo, A. Rajaraman and L. Susskind, Phys. Lett. {\bf B392} (1997) 319, [arXiv:hep-th/9605112].}
\REF{\HRS}{E. Halyo, B. Kol, A. Rajaraman and L. Susskind, Phys. Lett. {\bf B401} (1997) 15, [arXiv:hep-th/9609075].}
\REF{\EDI}{E. Halyo, Int. Journ. Mod. Phys. {\bf A14} (1999) 3831, [arXiv:hep-th/9610068]; Mod. Phys. Lett. {\bf A13} (1998), [arXiv:hep-th/9611175].}
\REF{\DES}{E. Halyo, [arXiv:hep-th/0107169].}
\REF{\UNI}{E. Halyo, JHEP {\bf 0112} (2001) 005, [arXiv:hep-th/0108167]; [arXiv:hep-th/0308166].}
\REF{\EXT}{E. Halyo, [arXiv:1506.05016].}
\REF{\CARL}{S. Carlip, Phys. Rev. Lett. {\bf 82} (1999) 2828, [arXiv:hep-th.9812013]; Class. Quant. Grav. {\bf 16} (1999) 3327,
[arXiv:gr-qc/9906126].}
\REF{\SOL}{S. Solodukhin, Phys. Lett. {\bf B454} (1999) 213, [arXiv:hep-th/9812056].}
\REF{\DL}{G. A. S. Dias and J. P. S. Lemos, Phys. Rev. {\bf D74} (2006) 044024, [arXiv:hep-th/0602144].}
\REF{\DGM}{S. Das, A. Ghosh and P. Mitra, Phys. Rev. {\bf D63} (2001) 024023,[arXiv:hep-th/0005108].}
\REF{\HSS}{M. Hotta, K. Sasaki and T. Sasaki, Class. Quant. Grav. {\bf 18} (2001) 1823, [arXiv:gr-qc/0011043].}
\REF{\MIP}{M. I. Park, Nucl. Phys. {\bf B634} (2002) 339, [arXiv:hep-th/0111224].}
\REF{\CARLI}{S. Carlip, Phys. Rev. Lett. {\bf 88} (2002) 241301, [arXiv:gr-qc//0203001].}
\REF{\DGW}{O. Dreyer, A. Ghosh and J. Wisniewski, Class. Quant. Grav. {\bf 18} (2001) 1929, [arXiv:hep-th/0101117].}
\REF{\CPP}{M. Cvitan, S. Pallua and P. Prester, Phys. Rev. {\bf D70} (2004) 084043, [arXiv:hep-th/0406186].} 
\REF{\KKP}{G. Kang, J. I. Koga and M. I. Park, Phys. Rev. {\bf D70} (2004) 024005, [arXiv:hep-th/0402113].}
\REF{\CHU}{H. Chung, Phys. Rev. {\bf D83} (2011) 084017, [arXiv:1011.0623].} 
\REF{\SIL}{S. Silva, Class. Quant. Grav. {\bf 19} (2002) 3947, [arXiv:hep-th/0204179].}
\REF{\MAJ}{B. R. Majhi and T. Padmanabhan, Phys. Rev. {\bf D85} (2012) [arXiv:1111.1809]; Phys. Rev. {\bf D86} (2012) 101501,
[arXiv:1204.1422].}
\REF{\EDIH}{E. Halyo, [arXiv:1406.5763].}
\REF{\EDIW}{E. Halyo. [arXiv:1403.2333].}
\REF{\WAL}{R. M. Wald, Phys. Rev. {\bf D48} (1993) 3427, [arXiv:gr-gc/9307038]; V. Iyer and R. M. Wald, Phys. Rev. {\bf D50} (1994) 846, [arXiv:gr-qc/9403028]; Phys. Rev. {\bf D52} (1995) 4430, [arXiv:gr-qc/9503052].}
\REF{\TRA}{D. Birmingham, K. S. Gupta and S. Sen, Phys.Lett. {\bf B505} (2001) 191, [arXiv:hep-th/0102051]; 
A. J. M. Medved, D. Martin and M. Visser, Phys.Rev. {\bf D70} (2004) 024009 [arXiv:gr-qc/0403026].} 
\REF{\CAR}{J. L. Cardy, Nucl. Phys. {\bf B463} (1986) 435.}
\REF{\FAT}{J. Maldacena and L. Susskind, Nucl. Phys. {\bf B475} (1996) 679, [arXiv:hep-th/9604042].}
\REF{\GIA}{A. Giacomini and N. Pinamonti, JHEP {\bf 0302} (2003) 014, [arXiv:gr-qc/0301038]; A. Giacomini,[arXiv:hep-th/0403183].}
\REF{\SEI}{N. Seiberg, Prog. Theo. Phys. Suppl. {\bf 102} (1990) 319.}
\REF{\LTY}{H. Erbin, Notes on 2d Quantum Gravity and Liouville Theory, \hfill \break
http://www.lpthe.jussieu.fr/~erbin/files/liouville{\_}theory.pdf.}
\REF{\ZAM}{A. Zamolodchikov and A. Zamolodchikov, Lectures on Liouville Theory and Matrix Models, http://qft.itp.ac.ru/ZZ.pdf.} 
\REF{\FRA}{Conformal Field Theory, P. Di Francesco, P. Mathieu and D. Senechal, Springer, 1997.} 
\REF{\CIG}{T. Mertens, H. Verschelde and V. I. Zakharov, JHEP {\bf 1403} (2014) 086, [arXiv:1307.3491];  
Phys.Rev. {\bf D91} (2015) 086002, [arXiv:1410.8009]; JHEP {\bf 1602} (2016) 041, [arXiv:1505.04025].}
\REF{\MAT}{I. Klebanov, [arXiv: hep-th/9108019]; P. Ginsparg and G. Moore, [arXiv: hep-th/9304011].}

\singlespace
\rl{SU-ITP-16/10}
\pagenumber=0
\normalspace
\medskip
\bigskip
\titlestyle{\bf{Liouville Theory on Horizons: Towards a Quantum Theory of Black Holes}}
\smallskip
\author{ Edi Halyo{\footnote*{e--mail address: halyo@stanford.edu}}}
\smallskip
\centerline {Department of Physics} 
\centerline{Stanford University} 
\centerline {Stanford, CA 94305}
\smallskip
\vskip 2 cm
\titlestyle{\bf ABSTRACT}
We show that any nonextremal black hole can be described by a Liouville theory that lives on its very near horizon region. In classical Liouville theory, the black hole corresponds to a field configuration that reproduces the Rindler metric. In quantum Liouville theory, the black hole state in the CFT is created by the puncture operator that gives rise to a conical singularity with a deficit angle of $2 \pi$. This state is ``heavy" with a nonnormalizable wave function in the minisuperpsace approximation. Black hole entropy counts the number of ways the ``heavy" black hole state can be obtained by acting with a product of ``light, quantum" operators in the CFT. Black hole hair is described by the ``background charge" in the CFT but its physical nature is unclear.

\singlespace
\vskip 0.5cm
\endpage
\normalspace

\centerline{\bf 1. Introduction}
\medskip

Despite the considerable progress in our understanding of black holes, the origin of nonextremal black hole entropy still remains mysterious. String theory can successfully account for the entropy of extremal and near--extremal black holes by using their BPS properties[\STR]; however the same methods cannot be applied to black holes far from extremality. The correct theory quantum gravity, which may or not be string theory in its different guises, is expected to explain the origin of nonextremal black hole entropy. Unfortunately, at the moment, it is not clear that we are close to such a deep and complete understanding of quantum gravity. In our present state of ignorance, it may be fruitful to search for a quantum mechanical description of nonextremal black holes and hope that this would provide important clues for the correct formulation of quantum gravity.

Recently, it was shown that any nonextremal black hole can be described by a state of a chiral, two dimensional conformal field theory (CFT) that lives in the very near horizon region of the black hole[\LAST]. The central charge of the CFT and the conformal weight of the state that corresponds to the black hole are determined by the dimensionless Rindler energy $E_R$. Then, the Wald entropy of the black hole is simply given by the Cardy formula. 
This method automatically applies to all nonextremal space--times which have near horizon regions described by Rindler space, including all nonextremal black holes and de Sitter space[\LEN-\UNI]. Moreover, extremal black holes can also be described by horizon CFTs in a special near horizon limit[\EXT].

In this paper, we concretely realize the horizon CFT idea in terms of a Liouville theory that lives in the very near horizon region of the black hole.
In order to investigate the near horizon physics of nonextremal black holes we dimensionally reduce the Einstein--Hilbert action
near the horizon by integrating out the transverse directions. The theory that arises from this procedure is two dimensional dilaton gravity with a computable dilaton potential. In the conformal gauge, the two dimensional metric is described by the only degree of freedom, i.e. the Liouville field. Thus, the near horizon physics is described by a coupled dilaton--Liouville field theory.
However, the constraint equations that arise from the stress--energy tensor show that the dilaton is fixed near the horizon. As a result, the only remaining dynamical degree of freedom is the Liouville field. In fact, the action we obtain is precisely that of Liouville theory that contains an exponential potential proportional to the cosmological constant and a linear coupling to the curvature scalar. 

In classical Liouville theory, the black hole is described by the field configuration that reproduces the Rindler metric. This is
not surprising since the very near horizon region of the black hole where Liouville theory lives is Rindler space. 
The solution we consider is the holomorphic half of the most general solution which also has a nonholomorphic part. We argue that the nonholomorphic part describes the past directed Killing horizon and is not physical. This leads to a chiral Liouville theory just like horizon CFTs. We then compute the central charge of the theory and the conformal weight of the solution and show that they precisely match those expected from horizon CFTs. As a result, using the Cardy formula, this solution correctly reproduces the black hole entropy. 

In quantum Liouville theory, the black hole state is created from the Liouville vacuum by the puncture operator with vanishing momentum. This operator introduces a delta function singularity in the curvature scalar. The black hole CFT state
is ``heavy" with a conformal weight that diverges in the classical limit. We argue that the same state can be created by a product
of ``light" or ``quantum" operators that vanish in the classical limit. Each such product that is compatible with the ``heavy" CFT state corresponds to a black hole microstate. There are $e^S$ different products of ``quantum" operators, where $S$ is given by the Cardy formula. The fully quantum states of the black hole are given by the different linear superpositions of these microstates. 
In the minisuperspace approximation to Liouville theory, the black hole wave function has vanishing momentum and is nonnormalizable.
Such states are local and microscopic, and play an important role in Liouville theory. 

This paper is organized as follows. In the next section, we review the horizon CFT description of nonextremal black holes. In section 3, we show that the very near horizon region of a nonextremal black hole is described by Liouville theory. In section 4, we describe
nonextremal black holes in classical Liouville theory. The quantum Liouville or the CFT description of black holes is given in section 5. In section 6, we show that our results can be generalized to $D>4$ in General Relativity. Section 7 contains a discussion of our results and our conclusions.

\bigskip
\centerline{\bf 2. Black Holes as Horizon CFTs}
\medskip

We begin by reviewing the horizon CFT description of nonextremal black holes.
{\footnote1{For previous work on CFTs that describe black holes see refs. [\CARL-\MAJ].}}
It is well-known that the near horizon geometry of these black holes in any theory of gravity is Rindler space.
Consider a black hole with a generic metric of the form
$$ ds^2=-f(r)~ dt^2+ f(r)^{-1} dr^2+ r^2 d \Omega^2_{D-2} \quad, \eqno(1)$$
in D--dimensions. The black hole radius, $r_h$, is determined by
$f(r_h)=0$. If in addition, $f^{\prime}(r_h) \not =0$, the black hole is nonextremal and the near horizon geometry is described by Rindler space. Near the horizon, $r=r_h +y$ with $y<<r_h$, which leads to the near horizon metric
$$ds^2=-f^{\prime}(r_h)y~ dt^2+(f^{\prime}(r_h)y)^{-1} dy^2+ r_h^2 d \Omega^2_{D-2} \quad. \eqno(2)$$
In terms of the proper radial distance, $\rho$, defined by $d\rho=dy/\sqrt{f^{\prime}(r_h)y}$  and the dimensionless Euclidean Rindler time $\tau=i(f^{\prime}(r_h)/2)~ t$, the near horizon metric becomes
{\footnote2{In the following all metrics will be in the Euclidean signature.}}
$$ds^2=\rho^2 d \tau^2 + d \rho^2 + r_h^2 d \Omega^2_{D-2} \quad, \eqno(3)$$
where the Rindler space, i.e. the metric in the $\tau$--$\rho$ directions naively looks like the flat metric in polar coordinates.

The dimensionless Rindler energy $E_R$ conjugate to $\tau$ is obtained from the Poisson bracket[\LEN]
{\footnote3{The i on the left--hand side is due to the Euclidean signature for time.}}
$$i=\{E_R,\tau\}=\left({\partial E_R \over \partial M}{\partial \tau \over \partial t}-{\partial E_R \over \partial t}
{\partial \tau \over \partial M} \right) \quad, \eqno(4)$$
where $M$ is the mass of the black hole conjugate to $t$. For classical black holes this leads to 
$$dE_R={2 \over f^{\prime}(r_h)}~ dM \quad.\eqno(5)$$
Using the definition of Hawking temperature, $T_H=f^{\prime}(r_h)/4 \pi$, we find that the black hole entropy is given by $S=2 \pi E_R$. 
This procedure can be used for all nonextremal black objects in any theory of gravity[\SBH-\UNI].
In fact, it can be shown that $E_R$, which is a holographic quantity that can be obtained from a surface integral over the horizon[\EDIH], is exactly Wald's Noether charge $Q$[\WAL] and therefore the Wald entropy of any nonextremal black hole is given by[\EDIW]
$$S_{Wald}=2 \pi Q=2 \pi E_R \quad. \eqno(6)$$ 

In ref. [\LAST], it was shown that any nonextremal black hole can be described by a state of a $D=2$ chiral CFT that lives in its very near horizon region. 
More specifically, the near horizon region of a nonextremal black hole is described by Rindler space with the metric in eq. (3), the dimensionless Rindler temperature $T_R=1/2 \pi$ and the dimensionless Rindler energy $E_R$. The physics in this region is described by a $D=2$ CFT since, in the vicinity of the origin of Rindler space, all dimensionful parameters are negligible and the transverse (to $\tau$ and $\rho$) directions decouple[\CARL,\SOL,\CHU,\TRA].
The horizon CFT is chiral since the Rindler metric in eq. (3) has only one $U(1)$ isometry which gets enhanced to a Virasoro algebra in the near horizon region. Therefore, we identify Rindler space with a chiral CFT with $L_0=E_R$. 
We also demand that the dimensionless Rindler temperature $T_R=1/2\pi$ be equal to the dimensionless CFT temperature, $T_{CFT}$, 
defined as
$$T_{CFT}={1 \over \pi} \sqrt{{{6 L_0^{\prime} \over c}}} \quad, \eqno(7)$$
where $L_0^{\prime}$ is the conformal weight of a state in Rindler space. Since (Euclidean) Rindler space
is obtained from the Euclidean plane by an exponential coordinate transformation, $L_0^{\prime}$ is shifted relative to the conformal weights of states on the plane[\LAST].

The entropy of the CFT state is given by the Cardy formula[\CAR]
$$S={\pi^2 \over 3}cT_{CFT} \quad. \eqno(8)$$
or equivalently by
$$S=2\pi \sqrt{{{c L_0^{\prime}} \over 6}} \quad. \eqno(9)$$
Setting $T_R=T_{CFT}=1/2 \pi$ we find that this shift in the conformal weights is $-c/24$ and thus $L_0^{\prime}=L_0-c/24$. 
As a result, the CFT state that describes the black hole satisfies[\LAST]
$$2L_0^{\prime}={c \over {12}}=E_R \quad. \eqno(10)$$
Substituting these values into the Cardy formula gives the correct Wald entropy in eq. (6).
Therefore, we can identify a nonextremal black hole with a $D=2$ CFT state that lives near the horizon and satisfies eq. (10). 

The Cardy formula is only valid asymptotically for $L_0^{\prime}>>c$ whereas for the CFT black hole state $L_0^{\prime}=c/24$. This problem is usually solved by invoking fractionation, i.e. by assuming that there are twisted sectors of the CFT[\FAT]. 
The dominant contribution to entropy comes from the most highly twisted sector with a twist of $E_R$. Due to the twist,
the central charge of the CFT and the conformal weight of the state become
$c=12$ and $L_0^{\prime}=E_R^2/2$ respectively. Now, the Cardy formula can safely be applied since $L_0^{\prime}>>c$, and
leads to the correct black hole entropy.

\bigskip
\centerline{\bf 3. Very Near Horizon Physics and Liouville Theory}
\medskip

In the previous section, we argued that the very near horizon region of a nonextremal black hole is described by a CFT
on Rindler space in the $\tau-\rho$ directions. In this section, we investigate the physics of this region and show that the horizon CFT described in the previous section is Liouville theory which is a theory of two dimensional quantum gravity. 
{\footnote4{This section is similar to ref. [\GIA]. However, the details and the interpretation are quite different since our starting point is the dimensionless Rindler space in eq. (3).}}


The (Euclidean) Einstein--Hilbert action for gravity in $D=4$ is given by 
{\footnote5{The generalization to higher dimensions is straightforward and is considered in section 6.}}
$$I_{EH}={1 \over {16 \pi G_4}} \int d^4x \sqrt{g} R(g) \quad, \eqno(11)$$
where $g_{\mu\nu}$ is the four dimensional metric. In order to dimensionally reduce this theory to one that lives in the $t-r$ 
directions we use the ansatz
$$ds^2=g_{ab} dx^a dx^b + {\Phi^2} d\Omega_2^2 \quad, \eqno(12)$$
where $g_{ab}$ is the two dimensional metric and $\Phi(x_a)$ is the field that parametrizes the radial direction. In the neighborhood of any $D=4$ nonextremal black hole with finite horizon radius $r_h$, we can dimensionally reduce the action by replacing 
$\Phi^2 \to r_h^2 \Phi^2$ in eq. (12). In this case $\Phi(x_a)$ becomes a dimensionless field that parametrizes the radial deviation from the horizon. Note that $\Phi(r_h)= 1$ fixed on or very near the horizon.
Dimensionally reducing the theory by using the ansatz in eq. (12), we obtain the two dimensional action
$$I={r_h^2 \over {4G_4}}  \int d^2x {\sqrt g} \left((2 \nabla \phi)^2+ \Phi^2 R(h) +2 \right) \quad. \eqno(13)$$
We now define the dilaton field, $\eta$ and rescale the metric by
$$\eta=\Phi^2 \qquad g_{ab}={1 \over \sqrt{\eta}}{h}_{ab} \quad, \eqno(14)$$
which leads to the two dimensional dilaton gravity action
$$I={r_h^2 \over {2 G_4}} \int d^2x \sqrt{h} \left({\eta \over 2}R(h)+ V(\eta) \right) \quad, \eqno(15)$$
where the dilaton potential is $V(\eta)=1/\sqrt{\eta}$.

In two dimensions, we can always choose to work in the conformal gauge in which the metric takes the form
$${h}_{ab}=e^{-2 \phi} \delta_{ab} \quad, \eqno(16)$$
where $\phi$ is the Liouville field which captures the two dimensional geometry and $\delta_{ab}$ is the Euclidean flat metric.
Using eq. (16) and the two dimensional curvature scalar $R=-2e^{-2 \rho}\partial^2 \phi$, eq. (15) becomes the action for the dilaton--Liouville theory that describes the physics near the black hole horizon
$$I={r_h^2 \over {2 G_4}} \int d^2x \left(-\partial_a \eta \partial^a \phi +V(\eta) e^{-2\phi} \right) \quad. \eqno(17)$$

Since Rindler space given by eq. (3) is flat, we can define complex coordinates on the plane, $z=X+iT$ and ${\bar z}=X-iT$, where
the flat coordinates $X$ and $T$ are related to those in eq. (3) by $X=\rho cos \tau$ and $T=\rho sin \tau$.
In complex coordinates, the equations of motions are given by
$$\partial_z\partial_{\bar z} \phi-{V^{\prime}(\eta) \over 4} e^{-2 \phi}=0 \qquad  \partial_z\partial_{\bar z}\eta+{V(\eta) \over 2} e^{-2 \rho}=0  \quad, \eqno(18)$$
whereas the constraints arising from the stress--energy tensor  
$$T_{zz}=T_{11}+T_{22} + 2T_{12}=0 \qquad T_{{\bar z} {\bar z}}=T_{11}+T_{22} - 2T_{12}=0 \quad, \eqno(19)$$ 
become  
$$\partial_z\partial_z \eta + 2 \partial_{z} \phi \partial_{z} \eta=0 \qquad \partial_{\bar z}\partial_{\bar z} \eta + 2\partial_{\bar z} \phi \partial_{\bar z} \eta=0 \quad. \eqno(20)$$
The mixed component of the stress--energy tensor vanishes on--shell, i.e. $T_{z {\bar z}}=0$, which means that, at least classically, the theory described by eq. (17) is conformally invariant.

The Euclidean Rindler metric which describes the geometry of the near horizon region of any nonextremal black hole can be written in the conformal gauge by using the complex coordinates $u^{\pm}=log \rho \pm i \tau$ (we remind that using the dimensionless Rindler time is equivalent to setting the surface gravity $\kappa=1$)
$$ds^2= exp(u^+ +u^-)~ du^+du^- \quad, \eqno(21)$$
from which we can read the classical Liouville field configuration
$$\phi=-{1 \over 2} (u^+ +u^-)= -log \rho \quad. \eqno(22)$$
We see that we can approach the horizon where $\phi \to \infty$ in two different ways: either by $u^+ \to - \infty$ or by
$u^- \to - \infty$. Nonextremal black holes have bifurcate Killing horizons which contain future directed and past directed Killing horizons which intersect over their event horizons. In the metric of eq. (21)
the past and future directed horizons are located at $u^+ \to - \infty$ and $u^- \to - \infty$ respectively. Since for a physical black hole there is only a future directed horizon, from now on we take $\phi=\phi(u^+)$ and ignore the $u^-$ dependence of the Liouville field. As a result, a physical nonextremal black hole is described by a chiral CFT. On the other hand, an eternal black hole which has both a future and past directed horizon would be described by a nonchiral CFT which depends on $u^-$ in addition to $u^+$. 

From the constraints in eq. (20) we find that
$$\partial_{z} \eta= exp(-2 \phi +\chi) \qquad \partial_{{\bar z}} \eta= exp(-2 \phi +{\bar \chi}) \quad, \eqno(23)$$
where $\chi=\chi({\bar z})$ and ${\bar \chi}={\bar \chi}(z)$. This means that the dilaton is constant near the horizon where $\phi \to \infty$. In fact, we know that $\eta(r_h)=\Phi^2(r_h)=1$ on the horizon so $\chi(z)={\bar \chi({\bar z})}=0$.
We conclude that, near the horizon, $\eta$ ceases to be a dynamical field and can be set to $\eta_h=1$. Therefore, the very near horizon region of the black hole is described only by the Liouville field $\phi$ that satisfies
$$\partial_z\partial_{\bar z} \phi-{V^{\prime}(\eta_h) \over 4} e^{-2 \phi}=0 \quad, \eqno(24)$$
where $V^{\prime}(\eta_h)=1/2 \eta_h^{3/2}$.

Eq. (24) for the Liouville field can be obtained from the Liouville action[\SEI,\LTY,\ZAM]
$$I=C \int d^2x \sqrt{h} \left({1 \over 2} {h}^{ab}\partial_a \phi \partial_b \phi +  \phi R(h) +\mu e^{-2 \phi}\right) \quad, \eqno(25)$$
where $\mu=-V^{\prime}(\eta_h)/2$ is the cosmological constant and $h_{ab}=\delta_{ab}$ is the flat metric in the $z,{\bar z}$
coordinates. The term with the curvature scalar does not contribute to the equation motion since Rindler space is flat. We will nevertheless keep it since it contributes to the stress--energy tensor and therefore to the central charge of the CFT.

In the following, we fix the constant $C$ in eq. (25) to be $C=r_h^2/G_4= 2 E_R$ so that it is consistent with eq. (15). The value of this constant is crucial since, as we will see below, it fixes the central charge of the Liouville theory and the conformal weight of the black hole state in the CFT.

The action in eq. (25) can be rewritten in terms of the rescaled Liouville field $\phi \to\sqrt{2E_R} \phi$ as
{\footnote6{The near horizon region of a black hole was also described by a Liouville theory in ref. [\SOL]. In that case the Liouville field was obtained by a field redefinition of the dilaton and did not arise from the two--dimensional metric.}}
$$I=\int d^2x \sqrt{h} \left({1 \over 2} h^{ab}\partial_a \phi \partial_b \phi + {1 \over b}\phi R(h) +\mu e^{-2 b \phi} \right) 
\quad, \eqno(26)$$
where $b=1/\sqrt{2E_R}$ and a factor of $b^2$ has been absorbed in $\mu$.
Therefore, the very near horizon physics of a nonextremal black hole is described by the Liouville theory which is a well--known
$D=2$ CFT with a calculable central charge and conformal weights.

\bigskip
\centerline{\bf 4. Black Holes in Classical Liouville Theory}
\medskip

In this section, we describe the nonextremal black hole in classical Liouville theory. We show that the black hole state is given by the classical Liouville field configuration that corresponds to Rindler space. This is not surprising since we found that both Rindler space and Liouville theory are descriptions of the black hole in the very near horizon region.

The generic action of Liouville theory, which describes two dimensional quantum gravity, is given by[\SEI,\LTY,\ZAM]
$$I_L=\int d^2x \sqrt{h} \left(h^{ab} \partial_a \phi \partial_b \phi +Q \phi R + \mu e^{-2 b\phi} \right) \quad, \eqno(27)$$
where the physical two dimensional metric in the conformal gauge is given by eq. (16).  This action arises from the two dimensional Einstein--Hilbert action with a cosmological constant $\mu$ in the conformal gauge given by eq. (16). The kinetic term for $\phi$ arises from quantization after taking into account the transformation of the path integral measure over metrics modulo diffeomorphisms. In our case, we see from eq. (17) that the kinetic term is due to the mixing between $\phi$ and $\eta$. The second term in eq. (27) arises from the coupling of the Liouville field to the conformal anomaly 
whereas the last term is due to the cosmological constant $\mu$. 

In the conformal gauge, the Liouville action in eq. (27), is invariant under the residual Weyl symmetry[\LTY]
$$h_{ab} \to e^{2 \alpha} h_{ab} \qquad \phi \to \phi+{\alpha \over b} \quad. \eqno(28)$$
Taking into account the rescaling $\phi \to b \phi$ in eq. (26), it is easy to see that the physical metric $g_{ab}$ is also invariant under eq. (28). The transformation of the Liouville field shows that it is the Goldstone boson of the Weyl symmetry which is spontaneously broken by the fiducial metric. Weyl symmetry is also broken explicitly by the cosmological constant $\mu$ and anomalously by the conformal anomaly. Under a conformal transformation $z \to w(z)$ the metric transforms as
$$h_{w {\bar w}} =\left |{dw \over dz} \right|^2 h_{z {\bar z}}  \quad, \eqno(29)$$
which is just a Weyl transformation of the type in eq. (28) with[\ZAM]
$$\alpha= log \left|{dw \over dz} \right| \quad. \eqno(30)$$
The Liouville field then transforms as
$$\phi \to \phi+{1 \over b} log \left|{dw \over dz} \right| \quad. \eqno(31)$$
The primary fields of the CFT are $e^{2a \phi}$ since
$$e^{2a \phi} \to \left|{dw \over dz} \right|^{2a/b} e^{2a \phi} \quad, \eqno(32)$$
under Weyl transformations.

It is easy to show that the classical Liouville theory described by eq. (27) is conformally invariant for $Q=1/b$[\SEI,\LTY,\ZAM].
{\footnote7{We are using the usual notation, $Q$, for the coefficient of the $R$ term in the Liouville action which has no relation to Wald's Noether charge in eq. (6).}} 
(Classical in the present context means that the theory has not been quantized as a path integral and the effects of this quantization, i.e. the integration measure over metrics modulo diffeomorphisms has not been taken into account.) 
We note that this is exactly the value in eq. (26), i.e. the action that describes the very near horizon geometry of nonextremal black holes. Thus, we conclude that nonextremal black holes are described by Liouville theory with $b=1/\sqrt{2E_R}$ which is, at least classically, a CFT. 

In the complex coordinates coordinates, $u^{\pm}$, the equation of motion obtained from the action in eq. (27) is
$$\partial_{+} \partial_{-} \phi + QR + \mu e^{-2 b\phi}=0 \quad, \eqno(33)$$
where $\partial_{\pm}$ denote derivatives with respect to $u^{\pm}$ respectively.
The second term vanishes since Rindler space is flat. The third term is negligible in the very near horizon region where 
$\phi \to \infty$. Thus, near the horizon, the Liouville field becomes a free scalar that is given by the sum of holomorphic and 
anti--holomorphic functions
$$\phi= \phi_1 (u^+)+ \phi_2({u^-}) \quad. \eqno(34)$$
This is a direct result of the fact that in the very near horizon region the exponential term in eq. (33) is negligible. Away from this region, the Liouville field is not free and there is a coupling between the holomorphic and anti--holomorphic parts.
As explained above, since we are interested only in the future directed horizon, we neglect $\phi_2$ and the $u^-$ dependence in eq. (34). The (holomorphic) stress energy tensor of the theory in eq. (20) is (in the $u^+$ coordinate)
$$T_{++}=\partial_{+} \phi \partial_{+} \phi +Q \partial_{+} \partial_{+} \phi \quad, \eqno(35)$$
where we took the near horizon limit, $\phi \to \infty$ which eliminated the usual term with the cosmological constant. 
$u^{+}$ is related to the flat complex coordinate $z$
on the plane by the exponential transformation $z=e^{u^+}$. Therefore, eq. (35) is related to the (holomorphic) stress--energy tensor in the flat coordinates by
$$T({u})d{u}^2=T(z)dz^2+{c \over 12}\{z,u\}dz^2 \quad, \eqno(36)$$
where 
$$\{z,u\}={z^{\prime \prime \prime} \over z^{\prime}}-{3 \over 2} \left(z^{\prime \prime} \over z^{\prime} \right)^2 \quad,\eqno(37)$$
is the Schwartzian derivative of the mapping. The coordinate transformation from $z$ to $u^+$ is exponential exactly
like the transformation from the plane to the cylinder, and thus we expect exactly the same shift in the stress--energy tensor. 

Since $u^+$ is periodic in the dimensionless Euclidean Rindler time direction, i.e. $\tau=\tau+2 \pi$, we can Fourier expand 
$T_{++}$ and obtain the components (for fixed $\rho$)
$$L_n={1 \over {2 \pi}} \int d\tau e^{-in \tau} T_{++}(\tau) \quad. \eqno(38)$$
The OPEs of the $L_n$ satisfy the classical Virasoro algebra with Poisson brackets[\SEI] 
$$i{\{L_n,L_m\}}_{PB} =(n-m)L_{n+m}+{c \over {12}}(n^3-n) \delta_{n+m,0} \quad, \eqno(39)$$
where the classical central charge is computed to be $c=6Q^2$.
We already found that $Q=1/b=\sqrt{2E_R}$. Therefore, the central charge of the theory is $c=6Q^2= 12 E_R$, i.e. precisely the value required by the horizon CFT.

We now proceed to find the conformal weight, $L_0^{\prime}$, of the CFT state that corresponds to the black hole. Since Liouville theory describes the near horizon region of the black hole, we assume that the classical Liouville field is given by eq. (22), i.e. it is the field configuration that corresponds to the two dimensional Rindler metric. Taking the rescaling $\phi \to b \phi$ used in 
eq. (26) into account, the Liouville field  solution becomes
$$\phi=-{\sqrt{E_R \over 2}} (u^+ +u^-) \quad. \eqno(40)$$
Using eq. (35) we find $T_{++}=E_R/2=Q^2/4$. Therefore, for the CFT state that corresponds to the black hole, 
$L_0^{\prime}=E_R/2$ which is precisely the value required by horizon CFTs. Note that we computed $L_0^{\prime}$ directly in the
$u^+$ coordinate rather than in the flat $z$ coordinate and taking the shift in eq. (36) into account. In the flat coordinates $L_0=0$ and the shift is exactly $E_R/2$ which gives the same result.

To summarize, we found that any $D=4$ nonextremal black hole can be described by a state in the two dimensional Liouville theory with
central charge $c=6Q^2$. Classically, the black hole state is described by the Liouville field configuration that reproduces  Rindler metric. This state has a conformal weight of $L_0^{\prime}=Q^2/4$. Using the Cardy formula in eq. (9), we find that this state has an entropy of $S=\pi Q^2=2 \pi E_R$ which is exactly the Wald entropy of the black hole. We conclude that the above Liouville theory is a concrete realization of the horizon CFT idea.

\bigskip
\centerline{\bf 5. Black Holes in Quantum Liouville Theory}
\medskip

In this section we consider a nonextremal black hole in the quantum Liouville theory. We obtain the CFT operator that creates the black hole state from the vacuum and find the black hole wave function in the minisuperpsace approximation. We begin
by canonically quantizing the theory described by the action in eq. (27) by Fourier expanding the field (in the coordinates of eq. (3))[\LTY]
$$\phi(\tau,\rho)= \phi_0(\rho) +\sum_{n \not=0} {i \over n} (a_n(\rho) e^{-i n \tau}+b_n(\rho) e^{in \tau}) \quad, \eqno(41)$$
and canonical momentum ($p={d \phi /d \tau}$)
$$p(\tau,\rho)= p_0(\rho)+ {1 \over {4 \pi}}\sum_{n \not=0} (a_n(\rho) e^{-i n \tau}+b_n(\rho) e^{in \tau}) \quad, \eqno(42)$$
in terms of the modes and creation and annihilation operators, where $a_{-n}=a_n^{\dagger}$ and $b_{-n}=b_n^{\dagger}$.
Note that in eqs. (41) and (42) the roles of the spatial and time directions are reversed compared to the usual conventions since in our case it is the Euclidean time direction that is periodic. Imposing the commutation relations
$$[\phi(\tau, \rho), p(\tau^{\prime}, \rho]= \delta(\tau-\tau^{\prime}) \quad, \eqno(43)$$
leads to
$$[\phi_0,p_0]=i \qquad [a_n(\rho),b_m(\rho)]=n\delta_{nm} \quad. \eqno(44)$$
If we substitute these into the stress--energy tensor in eq. (35) and demand that $T_{++}$ satisfy the Virasoro algebra and commute with $T_{--}$ we are lead again to an action of the type in eq. (27) in which the central charge and $Q$ are modified to[\SEI,\LTY]
$$c=1+6Q^2 \qquad  Q={1 \over b}+b \quad. \eqno(45)$$
Now, $b^2$ can be taken to be proportional to $\hbar$ that controls the quantum corrections[\LTY]. We already found that in the classical case $1/b=\sqrt{2 E_R} \sim \sqrt{S_{BH}}$ and
therefore $b<<1$ for large black holes. As expected, the quantum corrections to the classical Liouville theory are completely negligible for large black holes. Thus, the results of the previous section continue to hold in the quantum case and the black hole is described by a CFT state with $L_0^{\prime}=Q^2/4=E_R/2$ in Liouville theory with central charge $c=6Q^2=12E_R$. We note that the entropy of the black hole is inversely proportional to $\hbar$ as required, i.e. $S_{BH}=\pi Q^2= 2 \pi E_R/\hbar$.

Since Liouville theory is a CFT, it would be instructive to find the operator that creates the black hole state from the vacuum.
This would constitute a quantum description of the black hole or the quantum counterpart of the classical Liouville field in eq. (40) that corresponds to the black hole. In quantum Liouville theory, the primary states are created by the operators $V(z_0)=e^{2a\phi(z_0)}$ with
dimension $h_a=a(Q-a)$[\LTY,\ZAM]. The generalized momentum $a$ and the conformal weight of the state created by this operator are [\LTY,\ZAM] 
$$a={Q \over 2}+ip \qquad L_0^{\prime}={{Q^2} \over 4}+p^2 \quad. \eqno(46)$$
Since $Q=1/b=\sqrt{2E_R}$, the states with $p=0$ have weight $L_0^{\prime}=Q^2/4=E_R/2$ which, as we saw above, is precisely the weight of the the black hole CFT state. Therefore, the black hole state in quantum Liouville theory is given by
$$|black~ hole>=e^{Q \phi} |0>_L \quad, \eqno(47)$$
where $|0>_L$ is the Liouville vacuum. 

At this point a number of comments are in order. First, contrary to the usual CFTs, in Liouville theory there is not a one--to--one
correspondence between operators and states[\SEI]. This is due to the fact that the $SL(2,C)$ invariant vacuum state, $|0>_L$, is not in the
usual Hilbert space of the theory. However, the vacuum state can be created in a manner similar to a Hartle--Hawking state. It can
be obtained by a functional integral of the (minisuperpsace) wave function over a space with boundary with the insertion of the identity operator. Then, the black hole state in eq. (47) is obtained by a path integral over the same space with an insertion of the operator $e^{Q \phi}$. Thus, both
the vacuum and the black hole state exist in the Hartle--Hawking Hilbert space rather than the usual one for the CFT.

Second, the operator that creates the black hole state, $e^{Q \phi}$, is the puncture operator which creates a delta function
singularity in the Ricci scalar[\ZAM]. More precisely, the insertion of the operator $e^{2a \phi(z_0)}$ adds a term to the curvature scalar of the form
$$\sqrt{h} R(h)={8 \pi} {a \over Q}  \delta (z-z_0)  \quad, \eqno(48)$$
where $a=Q/2$ for the black hole state. We see that the puncture operator creates a point like singularity at $z_0$ with half the curvature of a sphere. This is a conical singularity with a deficit angle of $4 \pi a/Q= 2\pi$. Using the definition of $R$ 
and taking into account of the rescaling by $b$, $\phi$ satisfies the equation
$$\sqrt{h} R=-{8 \over Q} \partial_z \partial_{\bar z} \phi(z, {\bar z}) \quad, \eqno(49)$$
with the solution
$$\phi(z,{\bar z})=-{Q \over 2}log|(z-z_0)|^2 \quad. \eqno(50)$$
Using the relation between the $z,{\bar z}$ and $u^{\pm}$ coordinates, we conclude that the quantum field $\phi(z,{\bar z})$ should be identified with the classical Liouville field in eq. (40). The solutions in eqs. (40) and (50) match up to a shift of $z_0$ 
which is the point at which the puncture operator acts. Classically, $z_0$ is the location of the forward directed horizon. Thus the quantum field that corresponds to the puncture operator should be identified with the classical Liouville field that describes Rindler space. 

Third, since $b<<1$ for a classical black hole, $Q=1/b$ and the black hole state is given by 
$$|black~ hole>=e^{\phi/b} |0>_L \quad. \eqno(51)$$
This operator creates a ``heavy state" with a conformal weight that diverges in the classical limit, $b \to 0$[\LTY]. Its insertion contributes a delta function term to the equations of motion. On the other hand, if we take into account the quantum corrections, 
$Q=b+1/b$ and therefore the black hole state is created from the vacuum by a product of two operators
$$|black~ hole>=e^{b \phi}e^{\phi/b} |0>_L \quad. \eqno(52)$$
Here, the first operator creates a ``light state" with a finite conformal weight in the classical limit, $b \to 0$. Insertion
of this operator into the path integral does not contribute to the equations of motion. Thus, the fully quantum black hole state is given by a product of a ``light" and ``heavy" state in the classical limit.

Finally, the CFT spectrum that consists of primary operators of the type $e^{2a \phi}$ with dimension $h_a=a(Q-a)$ is symmetric under
the discrete symmetry $a \to Q-a$. It is interesting that the black hole state that is created by the operator with $a=Q/2$ is
self symmetric or corresponds to the fixed point of this symmetry.

The quantum picture of the black hole that emerges is as follows. The black hole state in the CFT is given by eq. (51) and
describes a ``heavy" state that is ``classical". In the present context ``classical" means that in the classical limit $b \to 0$
the conformal weight of the state diverges. Now, this ``heavy" state can be built from a product of ``quantum" or ``light" operators $e^{2n_i b \phi}$ with $L_0=n_i$ where $n_i$ are integers. (These are ``quantum" operators since they disappear, i.e. become the identity, in the classical limit.) Each such product of ``quantum" operators (acting on the vacuum) describes a quantum microstate of the black hole. Thus, we can write the black hole microstates as 
$$|black~ hole~ microstate >= \prod_{i=1}^{6Q^2} e^{2n_i b \phi} |0>_L   \quad, \eqno(53)$$
subject to the condition
$$\sum_{i=1}^{6Q^2} n_i={Q^2 \over 4} \quad. \eqno(54)$$
Above, the index $i$ runs over the different species of operators whose number is given by the central charge, $c=6Q^2$. The number of microstates is $e^S$ where the entropy $S$ is given by the Cardy formula in eq. (9).
It is mysterious that the Liouville field which is a two dimensional (though strongly self--interacting) scalar field can lead to such a large central charge corresponding to a large number of degrees of freedom.

The black hole quantum state in the CFT is given by the superposition of all the quantum microstates
$$|black~ hole>=\sum_{j=1}^{e^S} \alpha_j |black~ hole~ microstate>_j \quad, \eqno(55)$$
where $\alpha_j$ are complex coefficients and the index $j$ runs over all the microstates whose number is $e^S$. Eqs. (53)--(55) show the precise nature of the quantum black hole microstates and how they make up a black hole. The entropy of the ``heavy" black hole state with weight $Q^2/4$ arises from the different ways it can be built from the ``quantum" states with weights of order unity taking into account the different degrees of freedom given by the central charge of the CFT.

An alternative, more geometrical description of the black hole is based on the conical singularities created by the operators
$e^{2 b n_i \phi}$. We saw above that the quantum black hole state is created by the puncture operator $e^{Q \phi}$ which gives rise 
to a conical singularity with a deficit angle of $2 \pi$. Each operator $e^{2 bn_i \phi}$ creates a conical 
singularity with an infinitesimal deficit angle of $4 \pi bn_i/Q=4 \pi n_i/Q^2$. There are $6Q^2$ types of these singularities as
determined by the central charge.
Thus, we can create the conical singularity associated with the black hole state by the repeated application of these operators on the condition that the sum of the deficit angles add up to $2 \pi$, i.e. $4 \pi Q^{-2} \sum_i n_i=2 \pi$. This is equivalent to the condition in eq. (54) (up to a factor of 2). As a result, a black hole quantum microstate consists of a collection of a very large number of infinitesimal conical singularities that add up to a deficit angle of $2 \pi$. This picture is basically the geometric description of eqs. (53) and (54).

Even though we are able to exactly compute the Wald entropy of a black hole by using the Cardy formula in the Liouville CFT, the physical nature of the black hole hair is, unfortunately, not so clear. The central charge that counts the number of quantum black hole degrees of freedom is $c=1+6Q^2$ where $1$ is the quantum correction to the classical value $6Q^2$. In the Liouville Lagrangian, the parameter $Q$ appears in the term proportional to the curvature scalar which is necessary in order to take the
conformal anomaly into account. The classical Weyl symmetry of the theory requires $Q=1/b$. In the quantum theory, consistency of the CFT determines the values for the central charge and $Q$ as in eq. (45).

In a more physical picture of the CFT, the $Q\phi R$ term in the Lagrangian corresponds to an imaginary ``background charge" of $iQ$ at infinity[\FRA]. The Lagrangian of a scalar field in the presence of a ``background charge" $4 \pi q$ is 
$$I=\int d^2x \sqrt{h} \left({1 \over 2} h^{ab}\partial_a \phi \partial_b \phi + iq \phi R(h) \right) \quad. \eqno(56)$$
The (holomorphic) stress--energy tensor and central charge derived from above action are given by
$$T_{zz}=\partial_z \phi \partial_z \phi +iq \partial_z \partial_z \phi \qquad  c=1-6q^2 \quad. \eqno(57)$$

The free boson CFT has the shift symmetry, $\phi \to \phi+c$ which is broken by the second term in eq. (56). 
The current and the charge associated with this symmetry are defined as
$$j(z)=i \partial_z \phi \qquad \qquad  q={1 \over {2\pi i}} \int dz J(z) \quad. \eqno(58)$$
The charge $q$ is conserved because it commutes with $L_0$. A primary operator of the form $V_a=e^{2a \phi}$ carries a charge of $a$ since $[q,V]=aV$. In the presence of a ``background charge" of $q$, correlators such as $<V_{a_1}V_{a_2} \ldots V_{a_n}>$ are nonzero only if $\sum_i a_i-4 \pi iq=0$. If we interpret $q$ as a ``background charge", we see that the correlators vanish unless the total charge
of the system adds up to zero.

Comparing the above equations with eqs. (27), (33) and (45) we immediately conclude that Liouville theory has an imaginary ``background charge" of $q=-iQ$ at infinity. The neutrality condition for nonvanishing correlators now becomes 
$\sum_i a_i=4 \pi Q$[\ZAM].
The ``background charge" created by the $Q \phi R$ term in the Lagrangian can also be implemented by the operator $e^{2Q\phi}$ 
acting at infinity. This operator creates a conical
singularity with a deficit angle of $4 \pi$. Then, the neutrality condition simply states that, for nonvanishing correlators, the overall curvature has to add up to zero since Rindler space is flat.  

It is the ``background charge" $Q$ that is the origin of black hole hair. Due to the strong coupling nature of Liouville theory it seems that it is not possible to find weakly coupled degrees of freedom that correspond to hair. In fact, even though the central charge can be easily calculated from the OPE of the stress--energy tensor with itself, a microscopic counting of $c$ requires a better understanding of the physics associated with the ``background charge". 
Needless to say, the origin of black hole hair is a very important issue that requires further study.

We can also find the black hole wave function in the minisuperspace approximation in which we assume that the Liouville field is 
given by its zero mode, i.e. it is only a function of space $\phi(\tau,\rho)=\phi_0(\rho)$[\LTY]. In this case, the action becomes
$$S=\int d\rho \left({1 \over 2} {\dot \phi}^2 -2\pi \mu e^{-2 b \phi_0} \right) \quad, \eqno(59)$$
where the dot denotes the spatial ($\rho$) derivative. The Hamiltonian obtained from eq. (59) is 
$$H_0={1 \over 2} p_0^2 +2\pi \mu e^{-2 b \phi_0} \quad, \eqno(60)$$
where the momentum is $p_0={\dot \phi}$. The theory is quantized by defining
$p_0=-i{d/ {d \phi_0}}$.
It can be shown that the Schrodinger equation in minisuperspace is (using $H_0=L_0$)
$$H_0 \psi_p=\left(-{1 \over 2}{d^2 \over {d^2 \phi_0}}+2\pi \mu e^{-2 b \phi_0}\right)\psi_p=p^2 \psi_p \quad. \eqno(61)$$
Near the horizon the exponential term is negligible and we get the equation for a free particle with momentum $p$
$${d^2 \over {d^2 \phi_0}} \psi_p+ p^2 \psi_p=0 \quad. \eqno(62)$$
The solutions of eq. (62) with real $p$ are delta function normalizable and called nonlocal states. The solutions with imaginary $p$ diverge for $\phi_0 \to \infty$ and therefore are not normalizable. Nevertheless, these local states are not discarded since they play an important role in Liouville theory.

We already saw above that the black hole state created by the puncture operator has $p=0$. Therefore, the black hole minisuperspace wave function is given by 
$$\psi_0=c+d\phi_0 \quad, \eqno(63)$$ 
where $c$ and $d$ are constants. It is interesting that the black hole wave function is not normalizable; i.e. one of the states that one would naively discard. The wave function in eq. (63) should be matched with the one from the region in which $\phi_0$ is finite and the exponential potential is not negligible. This wave function is the solution to eq. (61) with $p=0$ and is given by the modified Bessel function.

The $p \not=0$ states are continuous deformations of the black hole state with $p=0$. Since $Q=1/b$ is a macroscopic quantity of
$O(\sqrt{S_{BH}})$ for $p<<Q$, these states have entropies that equal $S_{BH}$ up to very small corrections. Noting the CFT states that describe Rindler space satisfy $c=24 L_0^{\prime}$, we conclude that these states should correspond to deformations of Rindler space. One possibility is that the states with $p \not =0$ correspond to metrics of the form
$$ds^2=-(f^{\prime}(r_h)y+\epsilon y^2)~ dt^2+(f^{\prime}(r_h)y+\epsilon y^2)^{-1} dy^2+ r_h^2 d \Omega^2_{D-2} \quad, \eqno(64)$$
which is a one parameter deformation of the Rindler metric.
For small values of the parameter, i.e. $\epsilon y<<f^{\prime}(r_h)$, the space is very close to Rindler space with almost the same entropy.

We can consider Liouville theory that lives in the very near horizon region of the black hole in two different ways. First, we
can take it to be a pure CFT on flat space with a fixed metric. In this case, the central charge is a free parameter. In particular, it can be macroscopically large as above. However, the situation is different when Liouville theory is considered to be a theory of two dimensional quantum gravity with the dynamical metric $h_{ab}=e^{2 b \phi} \delta_{ab}$.
In this case, the consistency of the theory requires that the total central charge, $c_{tot}$, including the contributions from the Liouville field, matter and ghosts, add up to zero[\SEI,\LTY,\ZAM]. This is necessary for the unitarity of the theory which would otherwise be violated due to the conformal anomaly given by
$$<T^a_a>={{c_{tot}} \over {12}} R \quad. \eqno(65)$$
In our case, Liouville theory has a macroscopic central charge $c=6Q^2 \sim E_R \sim S_{BH}$ which cannot be canceled by the ghost contribution. However, we remind that, classically, the theory strictly lives on Rindler space which is flat and therefore $R=0$. Thus, we expect the classical description of the black hole to be valid even though the total central charge does not vanish. Quantum mechanically, the left--hand side of eq. (65) is the expectation value for the black hole state. On the right--hand side,
the puncture operator introduces a delta function singularity due to eq. (48). Therefore, $R$ is zero everywhere except at the point where the puncture operator acts, i.e. at the horizon. Clearly, whether this is sufficient
for the consistency of the quantum theory should be further investigated.

\bigskip
\centerline{\bf 6. Generalization to Higher Dimensions}
\medskip

The formula for Wald entropy $S_{Wald}=2 \pi E_R$ holds for any nonextremal black hole in any theory of gravity.
Above, we explicitly considered nonextremal black holes in four dimensional General Relativity. We now show that our results can easily be generalized to $D>4$ dimensions[\GIA]. In this paper, we do not attempt to show that our results also hold in generalized theories of gravity even though we expect that to be the case.

Consider the Einstein--Hilbert action in (Euclidean) $D$--dimensions
$$I_{EH}={1 \over {16 \pi G_D}} \int d^Dx \sqrt{g} R \quad. \eqno(66)$$
Since we will dimensionally reduce this $D$--dimensional theory by integrating out the angular directions near the horizon of a nonextremal black hole with radius $r_h$, we assume that the metric can be written as
$$ds^2=g_{ab} dx^a dx^b + {\Phi^2}r_h^2 d\Omega_{(D-2)}^2 \quad, \eqno(67)$$
where again $g_{ab}$ is the two dimensional metric and $\Phi(x_a)$ is the dimensionless field that represents the deviation from $r_h$ along the radial direction. 
Integrating out the angular directions we obtain the two dimensional action
$$\eqalignno{I={A_h \over {16 \pi G_D}}  \int d^2x {\sqrt g} [(D-2)(D-3)\Phi^{D-4}(\nabla \Phi)^2&+ \Phi^{D-2} R(h) &(68) \cr
&+(D-2)(D-3) \Phi^{D-4} ] \quad, \cr }$$
where $A_h$ is the area of the horizon. Defining the rescaled field $\Psi$ by
$$\Psi= C \Phi^{(D-2)/2}=\sqrt{{(D-3)} \over {2(D-2)}} \Phi^{(D-2)/2} \quad, \eqno(69)$$
we get
$$\eqalignno{I={{A_h} \over {4 \pi G_D}} \int d^2x \sqrt{g} [ {1 \over 2} (\nabla \Psi)^2&+ {1 \over 4} {{(D-2)} \over {(D-3)}} \Psi^2 R &(70) \cr
&+(D-2)(D-3) C^{-2(D-4)/(D-2)} \Psi^{2(D-4)/(D-2)} ] \quad, \cr }$$
where $A_h/4 \pi G_D=2E_R$. Defining the constants
$$a={{(D-2)} \over {(D-3)}} \qquad b=(D-2)(D-3) \left({1 \over {4C}}\right)^{2(D-4)/(D-2)} \quad, \eqno(71)$$
and redefining the dilaton and the metric as
$$\Psi^2=\eta \qquad g_{ab}={1 \over {a \sqrt{\eta}}} h_{ab} \quad, \eqno(72)$$  
the kinetic term is eliminated and the action becomes
$$I=2E_R \int d^2x \sqrt{h} \left({a \over 4} \eta R+  {b \over a} \eta^{(D-6)/2(D-2)} \right) \quad. \eqno(73)$$
Finally absorbing the constant $a/2$ into $\eta$ by a rescaling we find
$$I=2E_R \int d^2x \sqrt{h} \left({1 \over 2} \eta R+  V(\eta) \right) \quad, \eqno(74)$$
where
$$V(\eta)={b \over a} \left({{a \eta} \over 2} \right)^{(D-6)/2(D-2)} \quad. \eqno(75)$$
This action has the same form as the $D=4$ action given by eq. (15) with a dimension dependent dilaton potential. Thus, we can simply repeat the arguments in section 3
and all our results carry over to $D>4$ dimensions. We conclude that, in General Relativity, nonextremal black holes in all dimensions are described by classical Liouville theory as in section 4 or in quantum Liouville theory as in section 5.

\bigskip
\centerline{\bf 7. Conclusions and Discussion}
\medskip

In this paper, we provided a concrete realization of the horizon CFT description of nonextremal black holes. We found that the very near horizon geometry of black holes is described by Liouville theory with a calculable central charge. The black hole state in this CFT has a conformal weight that exactly reproduces the black hole entropy using the Cardy formula. More concretely, in classical Liouville theory, the field configuration that corresponds to the black hole reproduces the Rindler metric which describes the near horizon region. In quantum Liouville theory, the black hole state is created from the CFT vacuum by the puncture operator with zero momentum. The black hole state corresponds to a ``heavy" CFT state which can also be realized as a product of a large number of ``light", ``quantum" states. Each such product constitutes a quantum microstate of the black hole. The Cardy formula simply counts the logarithm of the number of these microstates. In the minisuperspace approximation, the black hole wave function is found to be nonnormalizable and describes a local and microscopic state. Black hole hair is described by the ``background charge" of the Liouville
theory that lives at infinity. Unfortunately, to the best oof our knowledge, the microscopic description of the ``background charge" in terms of weakly coupled degrees of freedom is not available. A better understanding of the ``background charge" is clearly crucial for a fundamental description of nonextremal black holes.

Liouville theory, which describes two dimensional quantum gravity, is closely related to two dimensional string theory[\MAT]. In our case, one can think of the dilaton and Liouville fields as the macroscopic dimensions of a target
space. In the very near horizon region, the dilaton is fixed, which means the theory actually lives on a one dimensional target space parametrized by the Liouville field. In addition, the exponential term that is proportional to the cosmological constant and leads to strong interactions in the world--sheet theory corresponds to a (massless) tachyon condensate. However, as we argued above, this term
is negligible in the very near horizon region and does not affect the physics. On the world--sheet, the term proportional to the curvature scalar corresponds to 
a linear (in the Liouville field which is a now space--time direction) dilaton vacuum. Since the string world--sheet 
lives on Rindler space which is flat this term vanishes as well. As a result, the string world--sheet theory describes one free scalar (or dimension) which is the Liouville field and another scalar which is not dynamical because it is fixed.
Therefore, our results can be interpreted in terms of two dimensional strings located in the the very near horizon region of the black hole, i.e. the two dimensional Rindler space. 
This is perhaps less surprising if we remember that in the 
very near horizon region all transverse modes are decoupled[\TRA] which should also apply to the transverse oscillations of the string. It is important to make this description more precise in order to describe nonextremal black holes in string theory.

There is another description of Rindler space in string theory that is based on the large radius limit of two dimensional stringy black holes[\CIG]. In this framework, Rindler space is described solely by the lightest mode of a long string near the horizon, i.e. the thermal scalar. It would be interesting to find out the relation between our results and those in refs. [\CIG].

The description of black holes in terms of Liouville theory on their horizons should also be able to reproduce Hawking radiation. 
In order to describe Hawking radiation, Liouville theory or the CFT must be coupled to outside degrees of freedom. The most straightforward way to do this is to couple the stress--energy tensor of the CFT to (outside) gravitons. The Liouville theory modes are at a finite dimensionless Rindler temperature of $T_R=1/2 \pi$ so their abundance is determined by the Boltzmann factor in a canonical ensemble. After a mode is emitted, black hole entropy decreases by one bit which can be realized in Liouville theory by acting on the black hole state in eq. (53) by the operator $e^{-2b \phi}$. It would be interesting to see if these preliminary ideas
can be used to reproduce the details of Hawking radiation. 

Finally, two dimensional quantum gravity or string theory can also be described in terms of the matrix model which is a discretized description of random surfaces[\MAT]. In a particular double scaling limit, this model reproduces Liouville theory which can be interpreted as criticality; i.e. that the more fundamental discrete matrix model flows to the continuous Liouville theory in the critical region.
It seems that our results can be better understood and perhaps generalized using the matrix model. This would be similar to understanding the Ising model from its CFT description at criticality. If criticality in this context corresponds to horizons, perhaps away from criticality we would be able to discover the discrete, fundamental building blocks of space--time.



\bigskip
\centerline{\bf Acknowledgments}

I would like to thank the Stanford Institute for Theoretical Physics for hospitality.

\vfill

\refout

\end
\bye